\documentclass{pasj01}

\RequirePackage{mathpazo}
\RequirePackage[T1]{fontenc}

\newcounter{author}
\def\authorcount#1#2{\refstepcounter{author}\label{#1}
                     \altaffiltext{\ref{#1}}{#2}}

\begin{document}
\SetRunningHead{T. Kato and N. Kojiguchi}{}

\Received{201X/XX/XX}
\Accepted{201X/XX/XX}

\title{BC Cassiopeiae: First Detection of IW And-Type Phenomenon Among
       Post-Eruption Novae}

\author{Taichi~\textsc{Kato},\altaffilmark{\ref{affil:Kyoto}*}
        Naoto~\textsc{Kojiguchi}\altaffilmark{\ref{affil:Kyoto}}
}

\authorcount{affil:Kyoto}{
     Department of Astronomy, Kyoto University, Kyoto 606-8502, Japan}
\email{$^*$tkato@kusastro.kyoto-u.ac.jp}



\maketitle

\begin{abstract}
IW And-type dwarf novae are recently recognized
group of cataclysmic variables which are characterized
by a sequence of brightening from a standstill-like phase
with damping oscillations often followed by a deep dip.
We found that the supposed classical nova BC Cas
which erupted in 1929 experienced a state of
an IW And-type dwarf nova in 2018, 89~yr
after the eruption.  This finding suggests that
high mass-transfer rate following the nova eruption
is associated with the IW And-type phenomenon.
The mass of the white dwarf inferred from the decline rate
of the nova is considerably higher than the average mass of
the white dwarfs in cataclysmic variables and the massive
white dwarfs may be responsible for the manifestation of
the IW And-type phenomenon.
\end{abstract}

\section{Introduction}\label{sec:intro}

   Cataclysmic variables (CVs) are composed of a white dwarf
and a mass-transferring secondary.  The transferred
matter forms an accretion disk around the white dwarf,
and in some cases, the accreted matter on the surface of
the white dwarf causes thermonucular runaway, which is
observed as a nova.  Dwarf novae (DNe) are a class of CVs,
which show outbursts resulting from instabilities
in the accretion disk
[for general information of CVs, novae and DNe
see e.g. \citet{war95book}].

   One of the long-standing questions about the relation
between novae and DNe is whether novae between eruptions
become detached systems (and hence not observed as CVs) or
become DNe.  This possibility was highlighted by
\citet{pat84CVevolution} after a comparison of the space densities
of novae and known CVs, and \citet{pat84CVevolution} suggested
that there are ``dead'' novae $\sim$100 times of CVs.
\citet{Hibernation} extended this discussion and suggested
the ``hibernation'' scenario by assuming that the mass-transfer
stops following angular momentum loss during a nova eruption
resulting in the binary to be detached.

   The relationship between novae and DNe have only recently
directly confirmed by observations: discoveries of
remnant nova shells around the DNe Z Cam \citep{sha07zcam}
and AT Cnc \citep{sha12atcncnova}, DN in AD 483 nova shell
Te 11 \citep{mis16Te11DN},
DN state before a nova eruption \citep{mro16hibernation}
and DN in AD 1437 nova shell \citep{sha17nsco1437}.

  Quite recently, a nova eruption
was recorded from the previously known DN V392 Per
by Y. Nakamura in 2018\footnote{
$<$http://www.cbat.eps.harvard.edu/unconf/\\
followups/J04432130+4721280.html$>$.
}  These observations indicate that at least some novae
erupt from DNe and some very old ($\gtsim$1000~yr old)
novae become DNe.

   In this paper, we report on the discovery of the DN
state of a recently recognized type with unusual characteristics
in the supposed classical nova BC Cas which erupted in 1929.

\section{BC Cassiopeiae}

   BC Cas was originally discovered as a long-period
variable by \citet{bel31bccas} on 1929 Simeis plates.
\citet{due84eyaqlbccasmtcenv745sco} studied Harvard plates
and presented a light curve.  The maximum brightness
was 10.7 mag (photographic) on 1929 August 1.
The object was not detected 14~d before.
Since the outburst amplitude was small and since
an old ``nova'' (HV Vir, 1929)
similarly recognized by the same author \citep{due84hvvir}
turned out to be a large-amplitude DN
(\cite{lei94hvvir}; \cite{kat01hvvir}),
BC Cas was suspected to be a candidate large-amplitude DN
and a search for a further outburst was conducted
[see a remark in \citet{kat01hvvir}].  This situation
continued until \citet{rin96oldnovaspec} published
a spectrum.  The spectrum showed relatively weak H$\alpha$
emission and a red continuum, which preferred a moderately
reddened nova rather (the object is indeed the direction
of a star-forming region SFR G115.80$-$1.6)
than a large-amplitude DN.
This classification has been confirmed by the small
Gaia parallax 0.490(71) mas and
the red ($Bp-Rp$=1.326) color \citep{GaiaDR2}.
According to \citet{sch18gaianova}, the distance
and Galactic extinction were estimated to be
2114($-$203,$+$557) pc and $A_V$=3.7, respectively.

\section{Zwicky Transient Facility Observations}\label{sec:obs}

\begin{figure*}
  \begin{center}
    \FigureFile(130mm,90mm){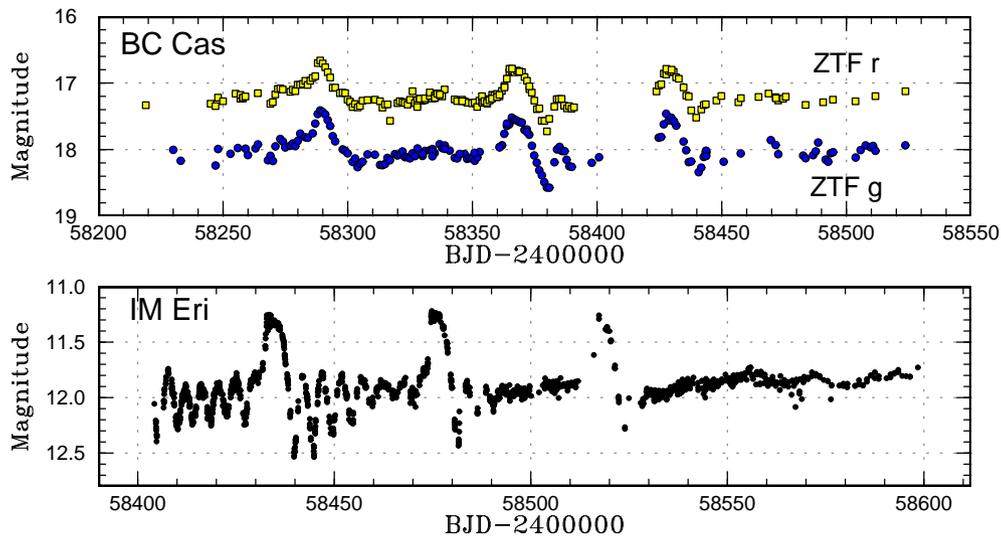}
  \end{center}
  \caption{(Upper) Light curve of BC Cas from ZTF observations.
  The filled squares and circles represent $r$ and $g$
  observations, respectively.
  (Lower) Light curve of IM Eri (typical IW And-type dwarf nova)
  for comparison.  The data were from \citet{kat20imeri}.
  }
  \label{fig:bccaslc}
\end{figure*}

   We used Public Data Release 3 of
the Zwicky Transient Facility (ZTF, \cite{ZTF})
observations.\footnote{
   The ZTF data can be obtained from IRSA
$<$https://irsa.ipac.caltech.edu/Missions/ztf.html$>$
using the interface
$<$https://irsa.ipac.caltech.edu/docs/\\
program\_interface/ztf\_api.html$>$
or using a wrapper of the above IRSA API
$<$https://github.com/MickaelRigault/ztfquery$>$.
}
We found that BC Cas showed DN-type outbursts
in 2018 (figure \ref{fig:bccaslc}).
In E-section 1, we also presented an analysis
of the reliability of the ZTF data in response to
the reviewer's opinion.  The ZTF data for BC Cas are
given in E-section 2.

   The three outbursts (BJD 2458288, 2458367 and 2458429)
were separated by 60--80~d and 
the brightness between the outbursts gradually increased
at least before the first and second outbursts.
Following the second outburst, the object showed
a deep dip.  The total amplitude of the outburst
including this dip was 0.9 mag.  There was also
a shallower dip after the third outburst.
Following the dip after the second outburst, there was
a hint of damping oscillation, though the details
were not sufficiently clear due to the observational
gaps.

\section{Discussion}

\subsection{Post-eruption BC Cas as an IW And-type dwarf nova}

   The features of the light curve of BC Cas 89-yr
after the (most likely) nova eruption in 1929 described
in section \ref{sec:obs} very well match the characteristics
of IW And-type DNe [see \citet{kat19iwandtype} and
references therein], which have been a recently recognized
type of DNe.\footnote{
   These objects are also called ``anomalous'' Z Cam stars
   \citep{szk13iwandv513cas}.
}
The IW And-type DNe show outbursts starting from
(quasi-)standstills and these outbursts are often followed
by dips and damping oscillations.  These DNe exhibit
such outbursts only a fraction of time and such a state
is referred to as IW And-type state or phenomenon.
The most remarkable feature is that this sequence is often
semi-periodically repeated.  The semi-regular occurrence
of this cycle suggests a yet unidentified type of limit-cycle
oscillation \citep{kat19iwandtype}.  BC~Cas is the first
(supposed) classical nova that experienced the IW And-type
state.

   The origin of the unusual variations of IW And-type
DNe is still poorly known.  \citet{szk13iwandv513cas}
suggested periodic increase of the mass-transfer from
the secondary.  \citet{kim20iwandmodel} suggested
that the tilted accretion disk could cause a limit-cycle
oscillation similar to IW And DNe.  The tilted disk,
however, has proven not to be a universal explanation
for IW And stars by the lack of negative superhumps,
a signature of a tilted disk, in many IW And stars
[e.g. IM Eri \citet{kat20imeri}].
Most recently, a detailed analysis of the Kepler data
of KIC 9406652, an IW And DN with prominent negative
superhumps, by \citet{kim20kic9406652} indicated
that neither models of \citet{szk13iwandv513cas} and
\citet{kim20iwandmodel} could explain the variation
of the disk radius.  The cause of the unusual pattern
of variation of IW And DNe still remains a mystery.

\subsection{Implications on IW And-type dwarf novae}

   In the hibernation scenario of classical novae,
the gradual reduction of the mass-transfer rate ($\dot{M}$)
from the secondary \citet{Hibernation}
enables the accretion disks in post-novae
to become thermally unstable to cause DN outbursts.
BC Cas showed the IW And-type phenomenon 89~yr after
the supposed nova eruption, which is one of the shortest
among classical novae to exhibit DN activity
(no other classical novae showed this phenomenon in
the ZTF data, and the case of BC Cas should be rare).
The condition, such as $\dot{M}$, enabling the IW And-type
phenomenon is still unknown.  Considering that BC Cas showed
this phenomenon during the cooling phase after the nova eruption,
it is likely that the IW And-type phenomenon in this object
was achieved in high-$\dot{M}$ condition probably close to
the border of thermal instability of the accretion
disk [see e.g. \citet{war95book}].  Whether this is
applicable to all IW And-type DNe requires further
study in other systems.

   According to \citet{due84eyaqlbccasmtcenv745sco},
BC Cas was a moderately fast nova with $t_3$
(time required to fade by 3 mag) of 50--75~d.
This translates to $t_2$ of 29--44~d using the formula
in \citet{hac06novadecline}.
It is widely accepted that nova eruptions occurring on
heavier white dwarfs have shorter $t_2$ or $t_3$
[see e.g. \citet{hac06novadecline}].  Using the model
grid in \citet{sha18novamass}, the mass of the white
dwarf in BC Cas is expected to be 1.02--1.09$M_{\odot}$.
\citet{zor11SDSSCVWDmass} obtained an average mass
of 0.82(3)$M_{\odot}$ for white dwarfs in CVs
with an intrinsic scatter of white dwarf masses of
0.15$M_{\odot}$.
A more recent study of eclipsing CVs yielded
a consistent result \citep{mca19DNeclipse}.
The inferred mass of the white dwarf in BC Cas
is more than 1$\sigma$ larger than the average
white dwarf mass in CVs and it may be that the IW And-type
phenomenon associated with the high mass of
the white dwarf.  This might explain why the IW And-type
phenomenon is seen only in limited number of DNe,
and why the same object repeatedly shows this phenomenon
while others never showed it.
Since no reliable estimates of white dwarf masses
are available in IW And-type DNe, more effort should
be paid to determine orbital parameters of these objects.

\section*{Acknowledgments}

Based on observations obtained with the Samuel Oschin 48-inch
Telescope at the Palomar Observatory as part of
the Zwicky Transient Facility project. ZTF is supported by
the National Science Foundation under Grant No. AST-1440341
and a collaboration including Caltech, IPAC, 
the Weizmann Institute for Science, the Oskar Klein Center
at Stockholm University, the University of Maryland,
the University of Washington, Deutsches Elektronen-Synchrotron
and Humboldt University, Los Alamos National Laboratories, 
the TANGO Consortium of Taiwan, the University of 
Wisconsin at Milwaukee, and Lawrence Berkeley National Laboratories.
Operations are conducted by COO, IPAC, and UW.

The ztfquery code was funded by the European Research Council
(ERC) under the European Union's Horizon 2020 research and 
innovation programme (grant agreement n$^{\circ}$759194
-- USNAC, PI: Rigault).

\section*{Supporting information}
Additional supporting information can be found in the online version
of this article.
Supplementary data is available at PASJ Journal online.

\end{document}